\pdfoutput=1

\documentclass[3p,10pt,twoside,twocolumn,number,sort&compress,final]{elsarticle}
\journal{Physics Letters A}

\usepackage{amsmath,amssymb}
\usepackage{stix,bm}
\usepackage{microtype}

\usepackage[colorlinks]{hyperref}
\usepackage{graphicx}

\newcommand{\ie}{i.{\kern1pt}e.}
\newcommand{\aka}{a.{\kern1pt}k.{\kern1pt}a.}
\newcommand{\rhs}{r.{\kern1pt}h.{\kern1pt}s.}
\newcommand{\iid}{i.{\kern1pt}i.{\kern1pt}d.}
\newcommand{\abs}[1]{\left|#1\right|}

\DeclareMathOperator{\EE}{\mathbb{E}}
\DeclareMathOperator{\var}{Var}
\DeclareMathOperator{\sgn}{\ensuremath{sgn}}

\makeatletter
\renewcommand\@makefntext[1]{\leftskip=0.0em\hskip-0.5em\@makefnmark{#1}}
\def\ps@pprintTitle{
   \let\@oddhead\@empty
   \let\@evenhead\@empty
   \def\@oddfoot{\footnotesize
       Submitted to Physics Letters A on May 28, 2020 \hfill Accepted for publication on July 20, 2020}
   \let\@evenfoot\@oddfoot}
\makeatother

\begin{document} \small

\begin{frontmatter}

\title{A numerical investigation into the scaling behavior of the longest increasing subsequences of the symmetric ultra-fat tailed random walk}

\author{J. Ricardo G. Mendon\c{c}a\href{https://orcid.org/0000-0002-5516-0568}{\includegraphics[viewport=-2 0 36 32, scale=0.25]{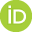}}}

\address{\centerline{Escola de Artes, Ci\^{e}ncias e Humanidades, Universidade de S\~{a}o Paulo, Rua Arlindo Bettio 1000, Vila Guaraciaba, 03828-000 S\~{a}o Paulo, SP, Brazil} \vspace{-18pt}}

\ead{jricardo@usp.br}

\begin{abstract}
The longest increasing subsequence (LIS) of a sequence of correlated random variables is a basic quantity with potential applications that has started to receive proper attention only recently. Here we investigate the behavior of the length of the LIS of the so-called symmetric ultra-fat tailed random walk, introduced earlier in an abstract setting in the mathematical literature. After explicit constructing the ultra-fat tailed random walk, we found numerically that the expected length $L_{n}$ of its LIS scales with the length $n$ of the walk like $\langle L_{n} \rangle \sim n^{0.716}$, indicating that, indeed, as far as the behavior of the LIS is concerned the ultra-fat tailed distribution can be thought of as equivalent to a very heavy tailed $\alpha$-stable distribution. We also found that the distribution of $L_{n}$ seems to be universal, in agreement with results obtained for other heavy tailed random walks.
\end{abstract}






\end{frontmatter}

\section{\label{intro}Introduction}

Let ${X} = (X_{1}, \ldots, X_{n})$ be a sequence of $n \geq 1$ real numbers and let $X_{i_{1}} < \cdots < X_{i_{k}}$, with $1 \leq i_{1} < \cdots < i_{k} \leq n$ be an (strictly) increasing subsequence of ${X}$ of maximum length $k$. We call any such subsequence a longest increasing subsequence (LIS) of ${X}$. There can be more than one LIS for a given sequence, with different elements but all of the same maximum length. While the LIS of random permutations (\aka\ Ulam's problem) has been the subject of much activity since the early 1970s \cite{ulam,hammersley,vershik,bdj99,patience,romik}, certainly due to its beautiful algebraic and combinatorial aspects, the fundamental properties of the LIS of random walks and other correlated time series remained little explored until recently \cite{angel,pemantle}, despite their applications in fields like data stream reliability and analytics \cite{liben,gopalan2007,gopalan2010,bonomi2016}.

In a couple of recent numerical studies, several properties of the LIS of random walks---its scaling behavior for different types of step length distributions, short and heavy tailed, large deviation rate function, and full distribution function---have been investigated \cite{lisjpa,hartmann,lispre}; see also the closely related \cite{howmany}. The behavior of the LIS of random walks with very heavy tailed distribution of step lengths, however, remained a challenge, mainly because of the difficulties involved in the efficient calculation with numbers of widely different orders of magnitude in the computer.

In this paper we consider the behavior of the length of the LIS of the symmetric ultra-fat tailed random walk introduced in \cite{pemantle}. In principle, we can think of an ultra-fat tailed distribution as equivalent to a symmetric $\alpha$-stable distribution with $\alpha \downarrow 0$. The expected length $L_{n}$ of the LIS of the ultra-fat tailed random walk has been shown to be bounded like
\begin{equation}
\label{eq:bounds}
n^{0.690} \leq \EE(L_{n}) \leq n^{0.815},
\end{equation}
but the exponents are not sharp \cite{pemantle}. Simulations of heavy tailed symmetric $\alpha$-stable random walks with $\alpha=1/2$ furnished $\EE(L_{n}) \sim n^{\theta}$ with $\theta \simeq 0.705$ \cite{lisjpa,lispre}.

In what follows, we simulate the symmetric ultra-fat tailed random walk and investigate the behavior of the length $L_{n}$ of its LIS as a function of the walk length $n$. Our goals are (i)~to verify whether the random walk specified in Section~\ref{sec:walk} indeed behaves like a very heavy tailed random walk vis-\`{a}-vis our previous results for such random walks and (ii)~how the behavior of its LIS compares with the bounds in (\ref{eq:bounds}). We also address the possible universality of the distribution function of the random variable $L_{n}$ for ultra-fat tailed random walks.


\section{\label{sec:walk}The ultra-fat tailed random walk}

In its original formulation, the symmetric ultra-fat tailed random walk involves a non-Archimedean totally ordered space that is not immediately suited for numerical approaches \cite{pemantle}. In \cite{limic}, the following characterization of an ultra-fat tailed distribution was given: in any collection of $n$ \iid\ picks from an ultra-fat tailed distribution, the greatest pick is much greater than the sum of the magnitudes of the other picks with probability tending exponentially rapidly to $1$ as $n \to \infty$.

Following \cite{pemantle,limic}, we can construct the symmetric ultra-fat tailed random walk for a fixed number $n$ of steps as follows. Let $b \geq 2$ be an integer, $\pi=(\pi_{1}, \ldots, \pi_{n})$ a random permutation on $\{1, \ldots, n\}$, and $s=(s_{1}, \ldots, s_{n})$ a sequence of $n$ independent $\pm$ signs picked uniformly at random. The $n$-step symmetric ultra-fat tailed random walk is the sequence $(X_{1}, \ldots, X_{n})$ given by
\begin{equation}
\label{eq:fat}
X_{0}=0, \quad X_{k} = X_{k-1}+s_{k}b^{\pi_{k}}, ~1 \leq k \leq n.
\end{equation}
The random signs ensure that the distribution of step increments is symmetric. The numbers $X_{k}$ can be represented by a vector with $k$ signs $s_{1}, \ldots, s_{k}$ located respectively at positions $\pi_{1}, \ldots, \pi_{k}$ and $n-k$ zeros elsewhere,
\begin{equation}
\label{eq:rep}
X_{k} = \big(s_{[\pi_{1}^{-1}\leq\,k]}, \ldots, s_{[\pi_{n}^{-1}\leq\,k]}\big),
\end{equation}
where by $s_{[\pi_{j}^{-1}\leq\,k]}$ we denote $s_{\pi_{j}^{-1}}$ if the inverse permutation $\pi_{j}^{-1} \leq k$ and $0$ otherwise. Equation (\ref{eq:rep}) is just the positional radix-$b$ representation of $X_{k}$.

The numbers $X_{k}$ are all distinct. Given $X_{i}$ and $X_{j}$ (assume for definiteness that $j>i$), we can check whether $X_{j} > X_{i}$ by checking
\begin{equation}
\label{eq:dif}
\sgn(X_{j}-X_{i}) = \sgn\bigg(\sum_{l=i+1}^{j}s_{l}b^{\pi_{l}}\bigg),
\end{equation}
which depends only on the sign of the highest power of $b$ on the \rhs\ We can thus determine whether $X_{j} > X_{i}$ by checking the sign $s_{k}$ with index $k$ determined by the condition $\pi_{k}=\max\{\pi_{i+1}, \ldots, \pi_{j}\}$. This means that we do not need to compute the numbers $X_{k}$ to compute the LIS of the sequence $(X_{1}, \ldots, X_{n})$, because the LIS of a sequence of numbers depends only on their relative order, not on their magnitudes. Given a realization of $\pi$ and $s$, we can compute the LIS of the random walk (\ref{eq:fat}) without ever having to actually write down the $X_{k}$.


\section{\label{sec:nums}Numerical results}

We generate the random signs $s$ trivially and the random permutations $\pi$ by the Fisher-Yates-Durstenfeld algorithm \cite{knuth}. Since we only sample an infinitesimal fraction of the possible $n!$ permutations in our simulations, we gloss over the fact that a pseudorandom number generator cannot produce more distinct permutations than it has distinct internal states. The LIS of an arbitrary sequence of numbers can be calculated in $O(n)$ space and $O(n\log{n})$ time by the patience sorting algorithm \cite{patience,sergei}.

\begin{figure}[t]
\centering
\includegraphics[viewport= 5 10 480 460, scale=0.225, clip]{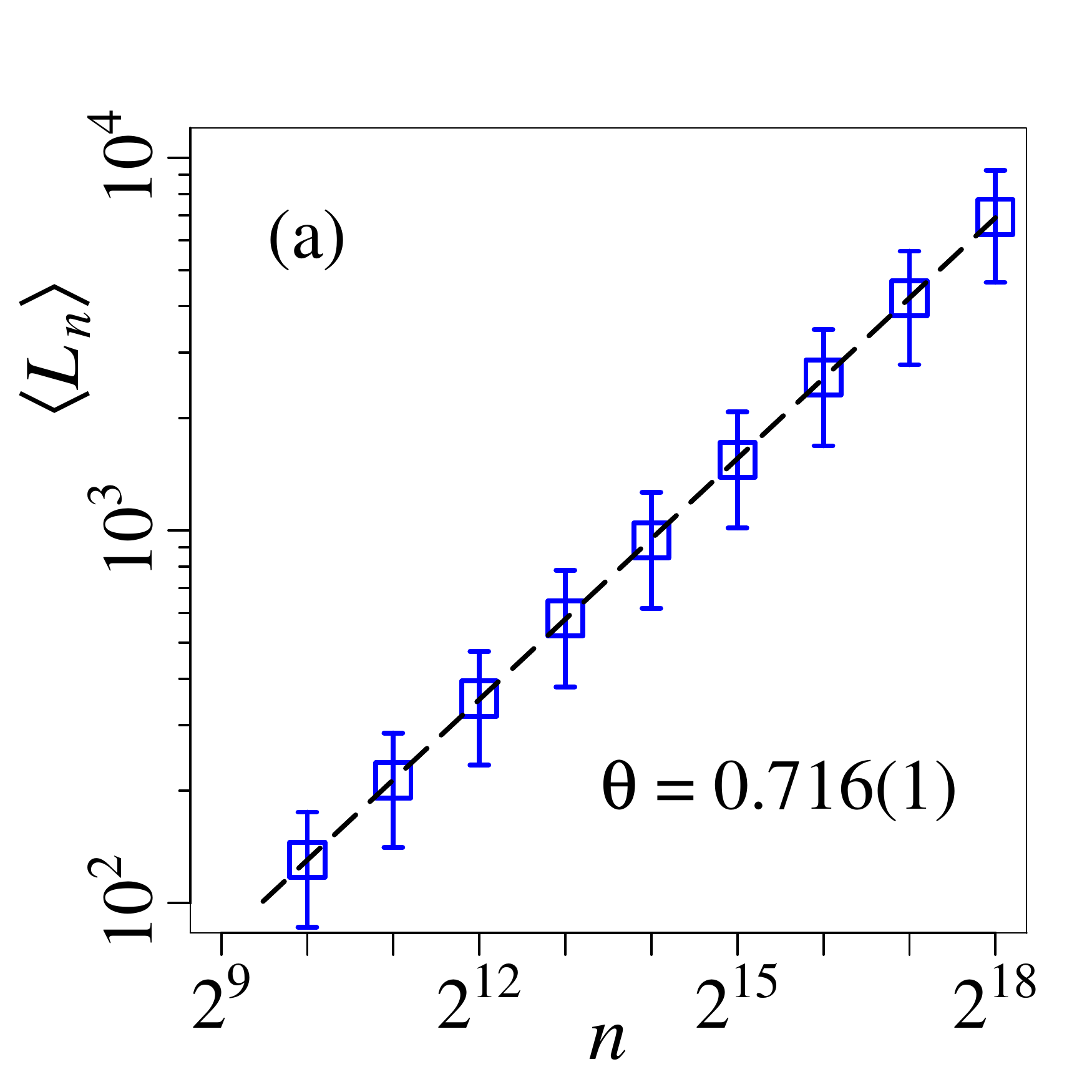} \hfill
\includegraphics[viewport= 0 10 480 460, scale=0.225, clip]{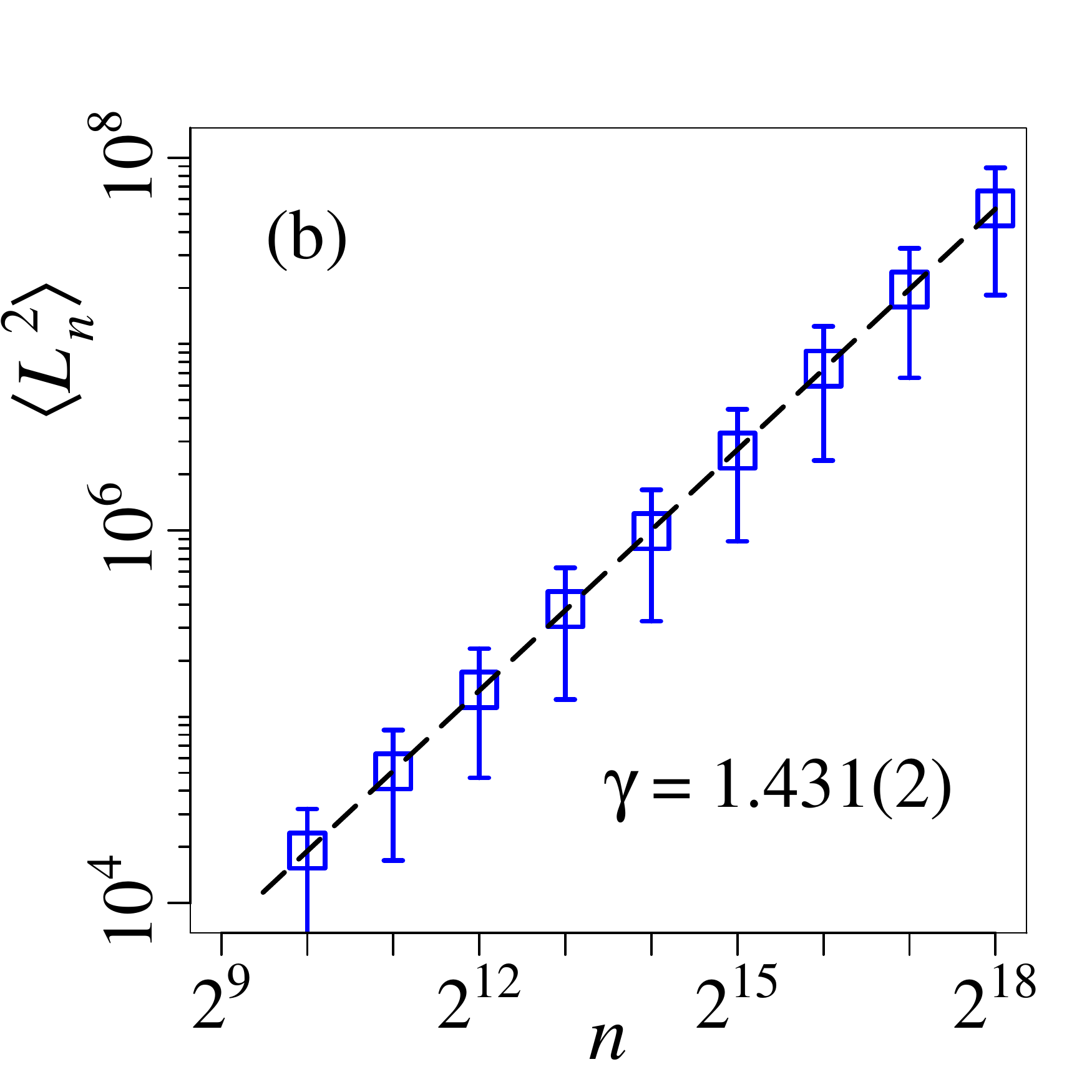}
\caption{Log-log plot of (a)~the sample mean and (b)~the sample second raw moment of $L_{n}$ over $5000$ sample $n$-step ultra-fat tailed random walks for each $2^{10} \leq n \leq 2^{18}$. The asymmetric error bars are an artifact of the logarithmic scales.}
\label{fig:lis}
\end{figure}

For each $n=2^{l}$, $10 \leq l \leq 18$, we generate $5000$ ultra-fat tailed random walks (\ie, pairs of random permutation and random signs), determine the length of their LIS and compute the sample raw moments $\langle L_{n} \rangle$ and $\langle L_{n}^{2} \rangle$. Our results appear in Figure~\ref{fig:lis}. A least-squares fit of the data to the function $an^{\theta}(1+b\log{n})$ furnishes
\begin{equation}
\label{eq:moms}
\begin{split}
\langle L_{n} \rangle \sim n^{\theta}, & \quad \theta = 0.716 \pm 0.002, \\
\langle L_{n}^{2} \rangle \sim n^{\gamma}, & \quad \gamma = 1.431 \pm 0.004.
\end{split}
\end{equation}
The best fits were obtained with $b=0$ in both cases

\begin{figure}[b]
\centering
\includegraphics[viewport=0 10 480 460, scale=0.30, clip]{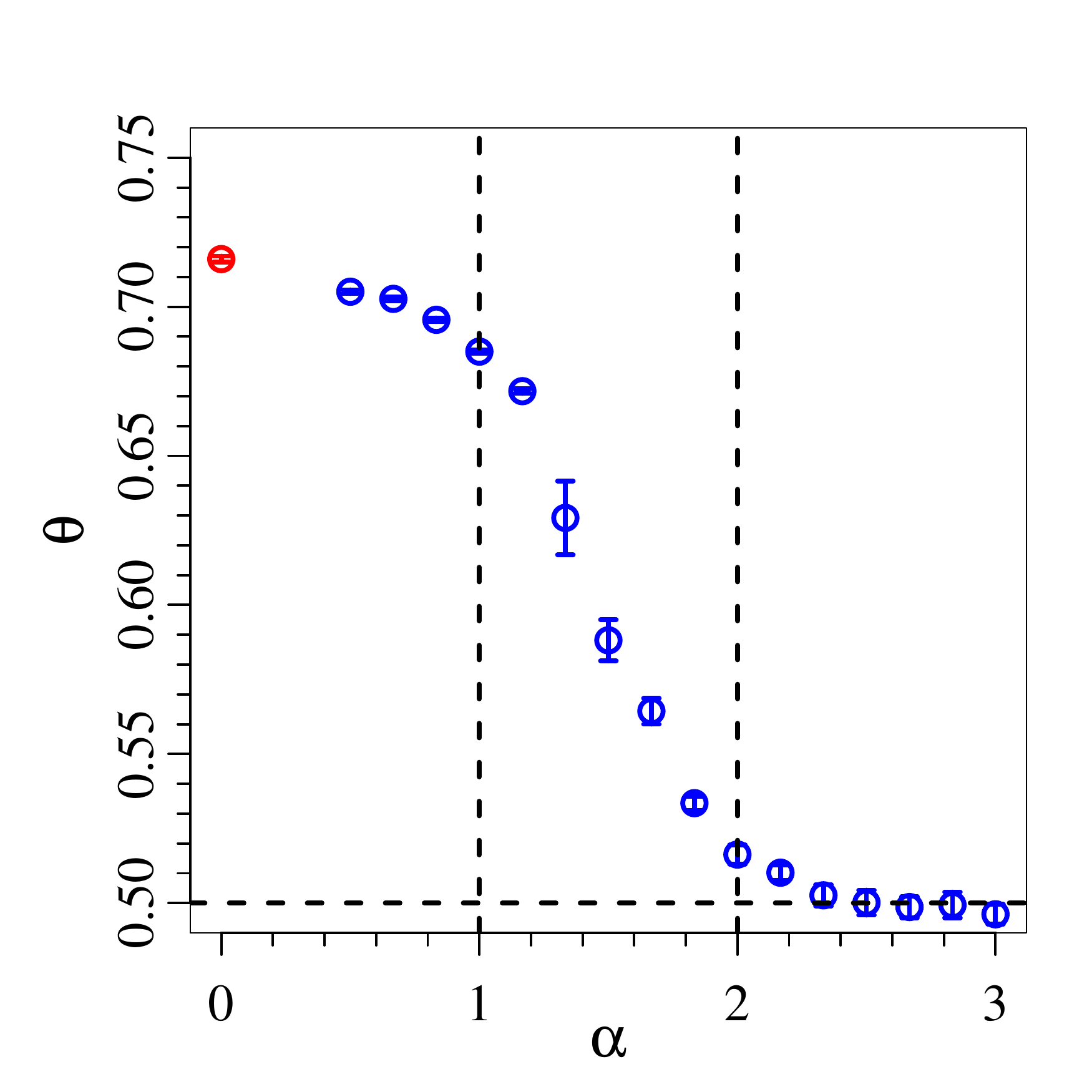}
\caption{Leading exponent $\theta$ in the asymptotic behavior of the LIS of random walks (see Eq.~(\ref{eq:moms})) with symmetric heavy tailed distribution of step increments $\phi(\abs{\xi} \gg 1) \sim \abs{\xi}^{-1-\alpha}$. The leftmost point is $\theta_{\text{ultra-fat}} = 0.716 \pm 0.001$, but its placement at $\alpha=0$ is disputable (see main text).}
\label{fig:theta}
\end{figure}

Figure~\ref{fig:theta} displays the exponent $\theta_{\text{ultra-fat}}$ together with the exponents found for random walks with symmetric heavy tailed distribution of step increments $\phi(\abs{\xi} \gg 1) \sim \abs{\xi}^{-1-\alpha}$ in the range $1/2 < \alpha \leq 3$. The data were taken from \cite{lispre}. We have placed $\theta_{\text{ultra-fat}}$ at $\alpha=0$ but this is disputable---while we know that the ultra-fat tailed distribution can be thought of as equivalent to a symmetric $\alpha$-stable distribution in the limit $\alpha \downarrow 0$ \cite[Thm.~6.1]{pemantle}, we do not know precisely to which $\alpha$ our finite-size ultra-fat tailed random walk corresponds. The fact that on a logarithmic scale $\alpha=0$ is infinitely far away from $\alpha=1/2$ does not help. We have not been able to find any simple functional relationship for $\theta(\alpha)$. We were initially expecting to find something like $\theta(\alpha) \sim (\alpha-\alpha_{c})^{z}$ for some $\alpha_{c}$, possibly equal to $1$ (below which $\EE(\xi)=\infty$) or $2$ (below which $\EE(\xi^{2})=\infty$), and ``critical exponent'' $z$, but the data do not support such scenario \cite{lispre}. From Figure~\ref{fig:theta} and our data we can only affirm that $\theta'(\alpha) \leq 0$ for $1/2 \leq \alpha \leq 3$ and that $\theta''(\alpha)$ changes sign from positive to negative somewhere between $\alpha=1$ and $\alpha=2$.

We see from (\ref{eq:moms}) that $\gamma$ is very nearly equal to $2\theta$. The same approximate relationship has been observed in other heavy tailed random walks \cite{lisjpa,lispre}. The fact that maybe $\gamma=2\theta$ exactly suggests that the distribution function of the random variable $L_{n}$ has the form
\begin{equation}
\label{eq:pdf}
f(L_{n}) = \frac{1}{\mathbb{E}(L_{n})}g\Big(\frac{L_{n}}{\mathbb{E}(L_{n})}\Big),
\end{equation}
for some distribution function $g(u)$. With such distribution, all moments of $L_{n}$ become proportional to $(\EE(L_{n}))^{k}$,
\begin{equation}
\label{eq:chgvar}
\EE(L^{k}) = \int\! L^{k}f(L) \mkern2mu dL = (\EE(L))^{k} \int\! u^{k}g(u) \mkern2mu du.
\end{equation}
We verified that, indeed, for our LIS data $\langle L_{n}^{3} \rangle \sim n^{2.998\theta}$ and $\langle L_{n}^{4} \rangle \sim n^{3.997\theta}$. The fact that (\ref{eq:pdf}) seems to hold for the LIS of all heavy tailed random walks (with distribution of step increments with infinite second and higher moments) with a universal function $g(u)$ irrespective of the underlying distribution of step increments was first noticed in \cite{lisjpa} and further verified in \cite{hartmann,lispre}. It has also been noticed that $L_{n}$ does not seem to be self-averaging in these cases, as $g(u)$ does not become more concentrated with increasing $n$. Recall that for a self-averaging random variable $\var(X) \to 0$ as $n \nearrow \infty$, as usual when the CLT applies, but many random variables encountered, e.\,g., in disordered systems, do not self-average, meaning that $g(u)$ approaches a non-Gaussian distribution, or, equivalently, that $L_{n}/\EE(L_{n}) \nrightarrow \text{constant}$ in distribution.

\begin{figure}[t]
\centering
\includegraphics[viewport=0 10 480 460, scale=0.30, clip]{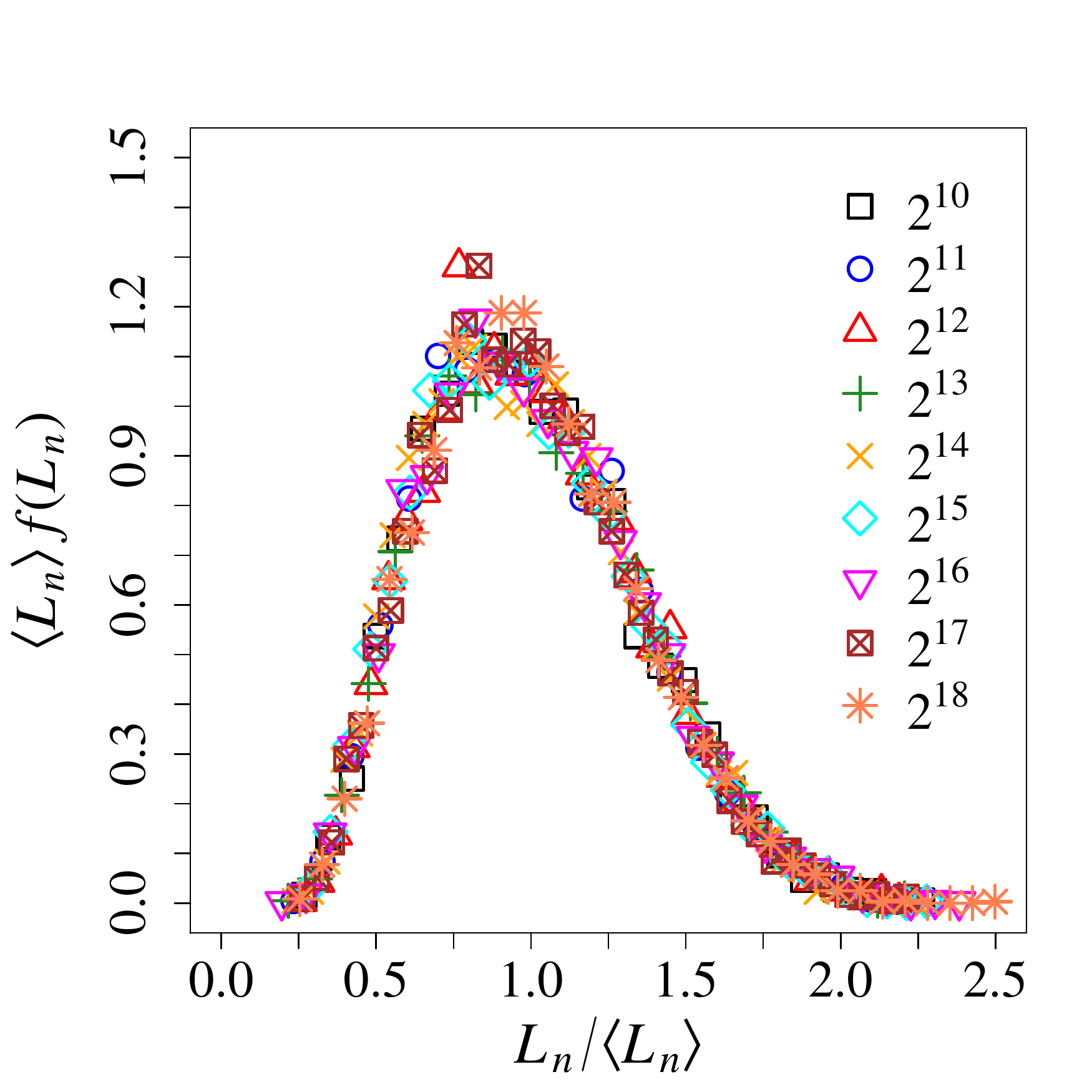}
\caption{Collapse of the $L_{n}$ data (binned into $24$ to $40$ bins, depending on the particular dataset) for the ultra-fat tailed random walk according to (\ref{eq:moms}) and (\ref{eq:pdf}). The ensuing curve is a sketch of~$g(u)$.}
\label{fig:scaling}
\end{figure}

Figure~\ref{fig:scaling} depicts $g(u)$ obtained from our data using expressions (\ref{eq:moms}) for $\EE(L_{n})$ and (\ref{eq:pdf}). The very good collapse observed strengthens the case for a universal distribution $g(u)$ for the length of the LIS of heavy tailed random walks, although we do not have a clue about its possible functional form. An attemp to relate $g(u)$ with a Gumbel distribution of number-theoretic pedigree proved unavailling, although the hypothetical connection between them could not be easily discarded either \cite{lispre}.

\section{\label{summary}Summary and conclusions}

We have extended previous studies on the length $L_{n}$ of the LIS of heavy tailed random walks by considering the symmetric ultra-fat tailed random walk, the step increments of which can be thought of as distributed according to a symmetric $\alpha$-stable distribution in the limit $\alpha \downarrow 0$. After explicitly constructing the ultra-fat tailed random walk, we found that $\mathbb{E}(L_{n}) \sim n^{\theta}$ with $\theta = 0.716 \pm 0.001$, which falls within the rigorous bounds (\ref{eq:bounds}) and also fits in nicely with the exponents found for other heavy tailed random walks, to wit, $\theta_{\text{ultra-fat}} > \theta_{\alpha=1/2}$, indicating that, indeed, as far as the behavior of the LIS is concerned the ultra-fat tailed distribution can be thought of as equivalent to a very heavy tailed $\alpha$-stable distribution with tail index $\alpha < 1/2$, at least \cite{pemantle,lisjpa,lispre}. The fact that $\theta$ is much closer to the lower bound than to the upper bound in (\ref{eq:bounds}) suggests that the upper bound might be improved, athough the rigorous analysis of the LIS of random walks seems to be a tough business. We verified that the scaling form (\ref{eq:pdf}) is obeyed by the LIS of ultra-fat tailed random walks, in agreement with results obtained for other heavy tailed random walks \cite{lisjpa,hartmann,lispre}. We currently do not know whether $g(u)$ corresponds to a known distribution. It may be possible, however, to employ the techniques developed in \cite{auffinger} (based on random recurrence relations) to tackle the min-plus tree model of Pemantle, a sort of stochastic coagulation-annihilation process moving inwards a binary tree towards the root node, to provide an approximate derivation for $g(u)$;  see also \cite{analco,derrida}. We hope to explore this possibility in the future.


\section*{\label{sec:ack}Acknowledgments}

The author thanks two anonymous reviewers for useful suggestions improving the manuscript and FAPESP (Brazil) for partial support under grants nos.~\mbox{2017/22166-9} and \mbox{2020/04475-7}.



\end{document}